# On the Radiation Effects of Strontium Ions on Satellite Solar Cells in Low Earth Orbits

Yue Chen[1], Miles A. Engel[1], Vania K. Jordanova[1], Misa M. Cowee[1], Gregory S. Cunningham[1], and David J. Larson[2]

[1]Los Alamos National Laboratory, Los Alamos, New Mexico, USA

[2]Lawrence Livermore National Laboratory, Livermore, California, USA

**Key Points:**

- High-altitude nuclear explosions generate strontium ions that pose radiation hazards to under-shielded low-Earth orbiting satellites

- Darkening effects on the satellite solar-cell coverglass are limited due to the fast decay and short ranges of strontium ions

- For exposed GaAs solar cells on satellites, strontium ions cause critical or even fatal degradation through displacement damages

**Abstract.** This study focuses on the radiation effects of $Sr^+$ ions—generated from high-altitude nuclear explosions (HANE)—on satellite solar cells in low-Earth orbits (LEO). Along four selected satellite orbits, ion fluences are sampled inside the evolving Sr+ distributions for days, determined from our newly developed HANE environment model. These fluences, along with the help of radiation transport codes including the MULASSIS and SRIM models, enable us to quantify the radiation damages by determining the values of total ionizing doses and the equivalent 1 MeV electron fluences for displacement damages. Comparing the dose values to existing experimental data, we conclude that HANE-generated $Sr^+$ ions have limited darkening effects to quartz solar cell coverglasses in LEO with apogees of 100s to 1000 km. In addition, with the extremely high equivalent fluences, we also conclude that these $Sr^+$ ions may cause severe or even fatal displacement damage to exposed solar photovoltaic (PV) cells on satellites in LEO. The radiation effects of $Sr^+$ ions are much less significant for the orbits with high apogees beyond ten thousand km. We also conducted model parameter sensitivity studies on the charge exchange cross-sections, neutral atmosphere density profiles and explosion local time positions, and the above conclusions stay unchanged. The methodology developed in this study can be extended to other HANE-generated heavy ion species in the future.

## 1. Introduction

With the first man-made satellite Sputnik successfully launched in 1957, the nuclear arms race of the Cold War quickly expanded to the new frontier of space. Starting from 1958, the USA and former USSR conducted a series of high-altitude nuclear explosion (HANE) tests. These tests not only significantly disturbed the atmosphere and ionosphere [e.g., Gregory, 1962], but also manifested unexpected new weapon effects compared to atmospheric nuclear tests, including such as powerful electromagnetic pulses and high levels of new radiation particles with destructive consequences in the global scale [e.g., Conrad et al., 2010 and Dupont, 2004].

In particular, several artificial mega-electron-Volt (MeV) electron (beta particle) belts were generated after the HANEs, and their observed sustaining existence poses a severe radiation threat to satellites [Hess, 1968]. For example, the Starfish Prime test, with a 1.4 Mt yield detonated at 400 km above Johnson Island in central Pacific Ocean on July 9, 1962, pumped the inner radiation belt with huge amounts of MeV electrons that have been well experimentally recorded [e.g., Dyal, 2006 and references therein] and trapped inside the geomagnetic field for a long time. Extensive observational and theoretical studies on Starfish electrons suggest this artificial belt has decay time scales ranging from days to years, depending on the electrons' location and energy [see references in Hess, 1968]. High fluxes of these MeV electrons are believed the culprit in knocking out and crippling a third of the satellites operating in low-Earth-orbits (LEO) following the Starfish Prime shot, including the failure of Navy satellites Transit 4B and TRAAC within several weeks due to deterioration of satellite solar power systems [Fischell, 1962]. Therefore, ever since the Starfish Prime test, understanding, quantifying, and curtailing the radiation effects of artificial radiation particles from HANEs have attracted a great deal of research interests for decades.

Besides the MeV electrons mainly sourced from the beta decay, a list of heavy ions are also produced as HANE fission fragments, including the radioactive strontium (Sr, with an atomic number of 38) isotopes that we consider here as a representative species for effects evaluations. Before turning into fallouts, these Sr particles are indeed in ionic forms and thus their movements are controlled by the geomagnetic field. Previous work by our modeling team at Los Alamos National Laboratory (LANL) updated the natural environment kinetic Ring current Atmosphere interactions Model (RAM) [Jordanova et al., 1997] with new HANE capability, by adding a localized ionized fission fragment source term based on high-fidelity first-principles models such as the KIM3D code [Larson et al. 2017] developed by Lawrence Livermore National Laboratory. In the initial runs of the RAM-HANE model, Engel et al. [2021] has showed how the radioactive $Sr^+$ ion cloud from a Starfish-like burst would follow the expected motions for ions in the inner magnetosphere, bouncing and drifting inside the Earth's dipole magnetic field, to form a distribution that extends away from the burst location and encircles the Earth. With this new RAM-HANE model, Engel et al. [2021] also investigated the $Sr^+$ ion spatial and energy distributions evolve with time by quantifying the loss and scattering effects of Coulomb collisions with the Earth's plasmasphere and charge exchange with the neutral atmosphere components. The former process smooths out the energy and pitch-angle distributions of $Sr^+$ ions, and the latter causes the removal of $Sr^+$ ions through neutralization and detrapping from the geomagnetic field.

Basically, this artificial Sr$^+$ ion belt is confined to low L-shells ($< \sim 2.5$) and has a strong dependence on particles' energy (up to 100s keV) and pitch angle. Readers are referred to Engel el al. [2021] for more details on the calculation, assumptions, and results of Sr$^+$ distributions.

This report specifically focuses on the radiation effects of Sr$^+$ ions on satellites. Starting with the results from Engel et al. [2021], the primary goal of this study is using radiation transport models to quantify how much damage these Sr$^+$ ions can cause to satellites orbiting in low altitudes. In the following, Section 2 first summaries how HANE generated Sr$^+$ ions are expected to interact with typical satellite solar panel system materials. Next, Section 3 presents the Sr$^+$ fluences anticipated for four selected LEOs. Then the darkening effects of Sr$^+$ ions on solar cell coverglasses are calculated for the four given orbits in Section 4, and also the displacement damages on solar photovoltaic (PV) cells are estimated in Section 5. We have some further discussions in Section 6, and the study is concluded in Section 7.

## 2. Stopping Power Functions and Ranges of Sr$^+$ Ions in Quartz

We first examined the stopping power curves of Sr$^+$ ions inside quartz (SiQ$_2$), which is the common main component for solar panel coverglasses. Figure 1a shows the Sr$^+$ ion linear energy transfer (LET) curve in red color for the electronic stopping power due to inelastic collisions with bound electrons in the medium, and the non-ionizing energy loss (NIEL) curve for the nuclear stopping power due to elastic collisions between Sr$^+$ ions and material atoms, both as a function ion energy. For comparison, the LET and NIEL curves for protons inside quartz are also plotted in blue color.  It is clear that Sr$^+$ ions have comparable LET values to protons except for the high energy end, but the NIEL values of Sr$^+$ ions can be up to more than ~30 times higher than their LET values, which is opposite to the case of protons. In addition, comparing to the MeV energy of HANE electrons, the energy range of Sr+ ions is mostly confined from ~ 0.1 to 100s keV, and here we use the 500 keV as the maximum (we will come back to this latter in Section 4).

By integrating the reciprocal stopping power functions over energy, one can determine the ranges of Sr$^+$ ions inside quartz, as shown in Figure 1b. The red curve shows that the Sr+ ranges increase with ion kinetic energy and have a maximum value of ~0.01 mil at 500 keV. In comparison, the blue range curve of protons have larger values due to their much smaller NIEL values. This max value of 0.01 mil is much lower than the typical thickness scales of solar panel system components, which are mostly at least 1 mil. The stopping power and ranges of Sr+ ions also have similar values (not shown here) inside GaAs (a common material of solar PV cells) and Aluminum (a material used for measuring satellite shielding). Hence, unlike the highly penetrating MeV electrons, HANE-generated Sr$^+$ ions cannot penetrate the surface layer of satellite components, and thus we expect Sr+ ions only affect the parts with no or very limited shielding/protection, such as the solar panels and optical lens.

The stopping power curves and ranges in Figure 1 are calculated using the 2013 version of Stopping and Range of Ions in Matter (SRIM-2013) code [Ziegler et al., 2008].This code implements the stopping power values of heavy ions as described in ICRU Report 73 [Bimbot et al., 2005]. The report also compares the stopping power values used in Geant4 code [Agostinelli et al., 2003] with those in SRIM code, and found almost identical values within the energy range of interest to this study. In addition, using the same SRIM-2013 code, we also derived the LET and NIEL curves and ranges for BoroSilicate glass (ICRU-169), an often used real coverglass material with minor compositions added to $SiO_2$ (e.g., B, Na and Al), and found small differences ($< 10\%$) when compared to those in Figure 1. Therefore, we proceeded using quartz ($SiO_2$) as the representative coverglass material with no loss of generality.

## 3. Sr$^+$ Ion Fluences and Dynamics along Satellite Orbits

Determination of Sr$^+$ ion radiation effects relies upon the temporally evolving distributions simulated by the RAM-HANE model as well as specified satellite orbits. Specifically, here we fly virtual satellites along selected orbits inside the ion distributions from Engel et al. [2021], to count the number of Sr$^+$ ions accumulated over time. Four satellite orbits are selected for this study, whose orbital parameters are summarized in Table 1. To compare how Sr$^+$ fluences vary with different orbits, here we select the International Space Station (ISS) to represent low-apogee orbits, two Transit satellites (TR4A and TR4B) to represent medium-apogee orbits with different inclinations, and the Telstar 2 (TS2) to represent high-apogee ones. (Appendix Figure A1 shows a top view of the orbits in equatorial plane.)

We first examine the details of Sr$^+$ ions observed along the ISS orbit. This orbit has an averaged altitude of ~400 km, inside of the Earth's thermosphere, and thus basically samples the low-altitude boundary of the Sr$^+$ cloud and encounter the ions before they hit their precipitation loss cone. Figure 2 presents the time-evolving fluences, over time bins with a size of 90 min, along ISS orbit during the first two days after the detonation. Panels A and B show the same energy differential fluences in different ways, in which a trough at ~10 keV is clearly visible in the first several time bins and is slowly smoothed out over time. From both panels, Sr$^+$ ions with energies above serval keV are seen to decay faster than lower energy ones due to Coulomb scattering with the plasmasphere. Panel C exhibits how the energy-integrated fluence for each time bin decays over the two days, showing the fluences decrease by more than six orders of magnitudes within the first 24 hr. Panel D plots the total fluence accumulated over the two days, which includes two energy components (below and above ~5 keV) and is dominated by the ions encountered in the first several time bins.

For the remaining three selected orbits, the orbits of TR4A and 4B have the majority of their time submerged inside the radiative Sr$^+$ cloud due to their higher apogees, while the TS2 orbit spends most of the time above the top of the radiative cloud. Figure 3 presents the Sr$^+$ ion fluences experienced in all four satellite orbits. Here a longer time bin size of 6 hr is used considering the orbital periods can be up to ~ 4 hr. The first four panels show the energy differential fluences for the four orbits decaying over six days after the detonation. One observation is that most Sr$^+$ ions dissipate before the end of day 4, and the higher the energy, the faster the decay. Ions observed by TS2 decays most slowly as in Panel D, probably due to relatively large pitch-angles of the sampled ions. Panel E confirms that the decay of Sr$^+$ ions in TR4A and 4B orbits is slower than for the ISS orbit and the decay speed in TS2 orbit is the slowest. In Panel F, the total fluence spectra by the end of day six are compared for different orbits, and the major difference between ISS, TR4A and TR4B orbits is for the high energy ($> $~10 keV) component, while the fluence in TS2 orbit has much lower values compared to the other three. Note the Sr$^+$ ions experienced in each orbit reflect the dependence on L-shell, pitch angle and local time of the simulated distributions in Engel et al. [2020] (e.g., Figures 2 and 3 therein). In general, orbits with lower altitudes sample Sr$^+$ ions interacting with more neutral atmosphere molecules, which thus have larger probabilities of being first removed through charge exchange.

Using the fluences presented in Figure 3, one can easily quantify the radiation effects using some radiation transport codes. The existing popular tools to analyze the effects of radiation environment on solar cells are the EQFLUX programme [Anspaugh, 1996 and references therein] developed by NASA Jet Propulsion Laboratory (JPL), and the SCREAM model [Messenger et al., 1999 and references therein] developed by US Naval Research Laboratory (NRL). The former calculates radiation damages based on determining the equivalent fluences compared to 1 MeV electrons, and the latter calculates the displacement damage dose from given incident particles. Both models are implemented in the SPENVIS website (https://www.spenvis.oma.be/), but only for the natural trapped particle populations. Therefore, we had to do our calculations in alternative ways. Next, we try to estimate and quantify Sr$^+$ ion radiation effects starting from the coverglass layer of solar panels on satellites.

## 4. Darkening of Solar Cell Coverglasses due to Ionizing Radiation from Sr$^+$ Ions

It has long been recognized that space radiation can cause the reduction of transmittance, or darkening, of solar cell coverglasses (and similarly of other cover coatings and adhesives) through the formation of color centers in glasses. When ionizing radiation excites orbital electrons from the valence band to the conduction band inside the cover materials, these electrons may be stably trapped by impurity atoms to form charged-defect complexes that absorb lights and thus reduce the transmittance [Carter and Tada, 1973; Anspaugh, 1996]. Per the solar cell handbook from Carter and Tada [1973], radiation effects in cover materials should be characterized by the ionizing damage, instead of displacement damage, which can be quantified by the absorbed total ionizing dose. Here we follow suit and calculate the Sr$^+$ ion ionizing dose (ID) on satellite solar cell coverglasses.

Considering the typical planar shape of solar panels, we used the Geant4 tool MUlti-LAyer Shielding SImulation Software (MULASSIS) for ID calculations. MULASSIS is a Monte-Carlo simulation-based tool that allows for dose and fluence analysis for selected incident particles through one-dimensional shielding with single or multiple layers [Lei et al., 2002]. The version of MULASSIS used for this study is the one implemented in the SPENVIS website. We adopt a two-layer geometry both made of quartz (SiQ$_2$): the top layer is the target layer for ID analysis with a thickness of 0.015 mil (i.e., 0.381 μm, about 1.5 times of the range of 500 keV Sr$^+$ ion as shown in Figure 1B), and the second layer has a thickness of 0.01 μm that is only used to confirm no Sr$^+$ ion making through the top layer.

The superposition principle has been applied to speed up dose calculations for this study. First, we determined the Sr$^+$ ion dose contribution function as shown in Figure 4 for the specific geometry described above. Each data point in this curve is obtained by feeding mono-energetic ions with a unit fluence level of 1/cm$^2$ to MULASSIS to calculate the corresponding ionization dose in the target layer. The derived curve is almost a power function with higher dose values for higher ion kinetic energies. Using this derived dose contribution function, one can quickly calculate $Sr^+$ ionization dose by integrating the product of this function and any given ion fluence spectrum over the whole energy range. The advantage of this superposition method is that it greatly shortens the calculation time when compared to starting a full-scale MULASSIS run each time. Precision of this method has been tested to have error percentages below 5% with one million incident ions, which is deemed adequate for this study.

We then proceeded to calculate how ID accumulate over the six days as well as the total ionization doses (TID) for the four orbits, using the energy-differential fluence spectra in Figure 3A-D as inputs. Here the TR4B orbit is used as the example to show the procedure step by step as well as the calculation results. As shown in Figure 5, taking the Sr+ ion fluence spectra in Panel A, we first calculate the ID value for each time bin, which decreases by order(s) of magnitude after every 6 hr (Panel B). The accumulated ID curve for TR4B plotted in Panel C shows the total dose is dominated by the IDs in day one, and the TID for TR4B by the end of day 6 has a value of 4.07 Mrad. Following the same approach, the accumulated ID curves for the other three orbits are also obtained and plotted in Panel C. Clearly, for all four orbits, the ionizing doses accumulated on day 1 dominate the TIDs. Plus, by the end of day six, the two Transit orbits have the highest TID of 7.57 and 4.07 Mrad, while the ISS orbit has a moderate dose of 2.16 Mrad and the TS2 orbit has the lowest value of 1.28 krad. This order of TID values is consistent with the total fluence spectra as shown in Figure 3F, which reflect how the radiative $Sr^+$ ion cloud being sampled by different orbits as discussed in Section 3.

It is useful to determine the effective energy range of $Sr^+$ ions in each orbit for this specific geometry. As in Figure 5D, the normalized accumulative dose curve for TR4B is plotted as a function of ion energy over the whole ion energy spectrum.  If defining the effective energy range as the core population of ions contributing to the TID from 5% to 95%, this range is ~ 3.5 – 40 keV for the Sr+ ion spectrum shape in the TR4B orbit as seen in Figure 3D. Of course, this effective energy range depends on the shape of an ion spectrum: the harder the spectrum, the higher the maximum energy value. For example, the high energy ends for the other three orbits increase to be from ~120 keV for TR4A, ~300 keV for ISS, and up to ~500 keV for TS2. Therefore, ion fluences outside these effective energy ranges—particularly ions with energy below multiple keV for the four orbits—can be safely ignored for TID calculations. This also explains why we may confine our study to the ions with energy up to 500 keV as in Section 2.

One open question is how reliable these calculated TID values are, or equivalently, how reliable those $Sr^+$ ion fluence spectra are calculated from RAM-HANE simulations in Engel et al. [2020].  Since unfortunately we have no experimental measurements to validate the calculated ion distributions, we chose to do a parameter sensitivity analysis on the charge exchange cross-sections (CECS). The CECS values used in Engel et al. [2020] were changed by applying multiplicative factors, to account for the factors such as uncertainties in theoretical calculation and/or the exosphere density profiles varying over solar cycle, and the RAM-HANE model was rerun with the changed CECS for new ion distributions. Again we use the TR4B orbit as the example with the results shown in Figure 6. Panel A replots the ion fluence distributions of using "normal" CECS as in Engel et al. [2020] for six days, which is identical to Figure 3C. Panel B (C) presents $Sr^+$ ion distributions with CECS decreased (increased) by a factor of 4, showing that decrease (increased) CECS values lead to slower (faster) decay of $Sr^+$ ions, particularly for high energies. Another way of understanding the effects of CECS is to compare the total ion fluence by the end of day 6, as in Panel D. It is seen that the changed cross sections have significant influence on the ions with energy above 1 keV, which falls in the effective energy range as determined above. Similar changes are seen for the other three orbits (not shown here). TID calculations were repeated for the four orbits with the new ion fluences generated in this step. Consequently, uncertainties in the CECS values lead to pronounced spreads in the final TID values.

All above calculated TIDs are plotted in Figure 7 as a function of the four orbits. The orange curve plots the TID with normal CECS, compared to the green one with reduced cross sections (and thus higher TID) and blue curve with increased cross sections (lower TID). The TS2 orbit always has the lowest dose values among the four orbits because of its high apogee. It is notable that the TS2 total dose increases from 1.28 krad in the normal case to 0.35 Mrad with the reduced CECS. For the rest three orbits, their TID values are within the range of ~2 – 4 Mrad for the normal case, and have CECS-caused variation factors ranging from ~4 (Transit 4A) to up to ~20 (ISS) depending on the orbits. In this plot, the highest TID is 9.44 Mrad for the ISS orbit with reduced CECS.

To quantify the darkening of quartz solar cell coverglass due to $Sr^+$ ionizing dose, we further compared the calculated TID listed in Figure 7 to the dose-transmittance loss curves from electron experiments—the only experimental curve we found in literature—as the Figure 3.23 in Carter and Tada [1973], assuming the same absorbed dose from either $Sr^+$ ions or electrons has the same darkening effects. We started with the worst scenario, i.e., the micro sheet coating. As in Figure 8, since the transmittance loss for the cell's maximum power curve is significant only for TID above 1 Mrad, here we focus the analysis on the three orbits with high TIDs. The three red dots mark the remaining transmittance corresponding to the TID values for three orbits (ISS, TR4A and TR4B) with normal CECS. The remaining transmittance values read from the curve are between 95.5% to 96.5%. For the highest TID of 9.44 Mrad for ISS orbit with reduced CECS (the rightmost green dot), the remaining transmittance is ~94%. In other word, the highest transmittance loss from our TID calculation is 6%.

However, the above transmittance loss values should be interpreted as the high limit. The transmittance-dose curve in Figure 8 is for highly-penetrating electrons with the assumption that the absorbed dose is almost even over the whole depth, which is not really the case for heavy ions such as $Sr^+$. Here we can do a quick estimate of the "real" remained transmittance. First, the whole coverglass depth of 0.0152 cm in Figure 8 can be divided into ~400 layers with an equal thickness of 0.381 μm. In the worst case of 94% remaining transmittance, with the incident MeV electrons, we should have $x^{400}$=94% where x is the "real" remaining transmittance after electron irradiation on each layer. Thus the "real" remaining transmittance has a value of 99.98% with ~10 Mrad dose from electrons. Since all $Sr^+$ ions deposit in the first layer with no effects on the other layers, it is reasonable to believe that the remaining transmittance loss due to Sr$^+$ is indeed 99.98%*1$^{399}$ = 99.98%, and thus the transmittance loss is much lower than 1% and is ignorable. In addition, the above analysis is for the micro sheet coating. If the coverglass is made of fused silica (i.e., the two top curves in Figure 8 for Corning 7940), the darkening effects are even smaller. Not to mention the recent technologies of such as Cerium-doped coverglass that is radiation-hardened or the latest new lightweight coating technologies [e.g., the POSS from Liu et al., 2008]. Therefore, we conclude that HANE-generated Sr$^+$ ions have limited radiation damages to solar cell coverglasses on satellites in LEO.

## 5. Displacement Damages from Sr$^+$ Ions on Exposed Satellite Solar Cells

Although it is unlikely for Sr+ ions to pass through the protective (including the coverglass) layer and hit the solar cells underneath, there is still a chance that some coating gaps exist on satellite solar cells. In this case, Sr$^+$ ions may cause significant degradation to the exposed solar cells, e.g., similar to the proton-caused sudden power losses on ATS-F1 and Intelsat II-F4 satellites [Carter and Tada, 1973 and Anspaugh, 1996]. Theoretically, when the incident energetic particles approach close enough to target nuclei, they can permanently displace some atoms from their original lattice sites. Then these displaced atoms and their associated vacancies undergo other reactions and finally form stable defects, which significantly decrease the minority carrier lifetime and diffusion length and thus reduce the power of a solar cell. Therefore, we need to calculate the displacement damage from Sr$^+$ ions so as to quantify their effects on a solar cell.

Here we follow the procedure of Messenger et al. [2005] that is to use the SRIM model to determine the relative damage coefficients (RDCs) of Sr$^+$ ions on a GaAs solar cell. As mentioned in Section 3, there are two general ways of quantifying displacement effects on solar cells: One is to calculate RDCs that is the technical route adopted by JPL's EQFLUX model, and the other is to calculate the displacement damage dose (based on the NIEL functions) that is the approach used by NRL's SCREAM model. The two models usually give similar results on GaAs/Ge solar cells [Messenger et al., 2001]; however, as pointed out by Messenger et al. [2005], the NIEL approach works well only for the particle energy that can be assumed to be almost constant across the active region of a solar device. This is generally the case for MeV electrons and protons but not for the Sr$^+$ ions in this study (see the range plot in Figure 1B). Particularly, Messenger et al. [2005] points out that the NIEL approach may not give satisfying results for calculating the RDCs for proton energies <100 keV, mainly because the "slowing-down" effects of incident protons deposited inside the device can cause maximum damage when proton tracks terminate inside the active device region. Therefore, as described in Messenger et al. [2005], our calculation involves using SRIM model to determine the vacancies generated inside a GaAs cell, and then normalize the results to the number of total vacancies produced by 10 MeV protons to obtain the equivalent proton fluences, and finally to get the equivalent 1 MeV electron fluences so as to compare to experimental results.

Once again, the same superposition method is used. We first determined the Sr$^+$ ion vacancy contribution function as shown in Figure 9 for the specific solar cell geometry: a layer of GaAs slab with a thickness of 0.025 mil. Each data point in this curve is obtained by feeding mono-energetic ions with a unit fluence level of 1/cm$^2$ to SRIM to calculate the total vacancies generated inside in the target cell. The derived curve is almost a power function with more vacancies for higher ion kinetic energies. Using this derived vacancy contribution function, one can quickly calculate the total vacancies by integrating the product of this function and any given ion fluence spectrum over the whole energy range. For comparison, 10 MeV protons with the unit fluence generate 0.03 vacancies inside the cell.

Using the same ion fluences as in Section 4, the total vacancies for all four orbits are calculated and plotted in Figure 10: here the orange curve plots the vacancies with normal CECS, the green one is with reduced cross sections (and thus higher fluences and more vacancies) and the blue curve with increased cross sections (lower vacancies). All curves have the similar shape as those in Figure 7. Again the TS2 orbit always has the least displacement damages. For the other three orbits, with the normal CECS the total vacancies are from $1.28 – 4.21 \times 10^{15}$, with variation factors ranging from ~4 to up to 20. In this plot, the highest vacancy number is $4.67 \times 10^{16}$ for the ISS orbit with reduced CECS.

Divided by 0.03, the above total vacancies can be converted to equivalent 10 MeV proton fluences; then multiplying by another factor of 1000, one can further convert them to the equivalent fluences for 1 MeV electrons on GaAs cells [Anspaugh, 1996]. That is, the equivalent 1 MeV electron fluences are simply the total vacancies multiplied by $3.33 \times 10^4$. These equivalent fluences can be compared to the experiment results, as in Figure 11 (copied from the three curves from Fig 6.29 from Anspaugh [1996]. Note the dots in the plot have their equivalent electron fluences reduced by a factor of $3.33 \times 10^3$ so that they can stay inside the plotting window (e.g., the real fluences for the three red dots should have values between $10^{19} – 10^{20}$ and thus the max power will be < 0.1). Therefore, based on the three power-fluence curves in Figure 11, exposed GaAs cells in the ISS, TR4A and TR4B orbits may not survive the displacement damages from Sr+ ions. The only exception is the TS2 orbit which may have acceptable losses in power for the normal and increased CECS cases.

Surely we still need to be careful with the above results since MeV electrons can penetrate the cells in Figure 11 but not Sr$^+$ ions. Indeed, the displacement damage to an exposed solar cell depends on its specific design. Although the penetration ranges of Sr+ ions are limited, they could lead to a total failure of the whole cell junction. For example, for devices with the space charge region 0.4 -1 μm below the cell surface, low energy protons (10s -100s keV) over a narrow (2 mil) gap can cause junction damages that were sufficient to drastically alter the cell device's power generating capability [Carter and Tada, 1973 and Anspaugh, 1996]. It is reasonable to expect the same for Sr$^+$ ions. Considering the ~0.2 μm range of 500 keV Sr+ ions inside GaAs, a severe failure of the ASEC cell in Figure 11 is still possible, which has a window thickness of 0.1 μm and junction depth of 0.45 μm; for the HRL cell with a window thickness of 0.3 μm, however, a severe junction damage is unlikely and the damage may confine to the top layer of the exposed portion. Of course, going into the specific cell design is beyond the scope of this study. Therefore, we conclude that HANE-generated Sr$^+$ ions may have severe or even fatal displacement damages to exposed solar PV cells on satellites in LEO.

## 6. Discussions

We repeated the RAM-HANE simulations to test how sensitive the above results are to input parameters. Here we call the Sr$^+$ ion distributions used in previous Sections the "normal" case which has the explosion at midnight and atmosphere densities as the solid lines in Appendix Figure A2. We also tested two other cases with the same modified atmosphere densities (dashed lines in Figure A2): One has the explosion moved to the noon (MLT = 12, called Case 1) and the other stays at midnight (MLT=0, called Case 2). Sr$^+$ ion fluences experienced in the same four orbits for these three cases are compared in Figures 12 – 16. Here only show distributions in the first two days because of their dominance to the total fluence/radiation.

In Figure 12, Sr$^+$ ions in the first time bin for Case 1 (Panel C) clearly have higher intensities at > 10s keV, while the decay speed of 10s keV in Case 2 (Panel B) is faster than the others. By the end of day 2, the total ion fluence in Case 1 is the highest, followed by the normal case and Case 2 is the minimum. Things are different for Transit 4A in Figure 13. It is interesting to see in the Panel D that Sr$^+$ ion fluences in the normal case is the highest while that from Case 1 is the lowest. For Transit 4B in Figure 14, Sr$^+$ ion fluences in both Case 1 and Case 2 have similar values, except for >100 keV energies, and both are lower than those from the normal case. For the Telstar 2 orbit, ion fluences from Case 1 is the highest as in Panel D, while those from Case 2 is the lowest.

Using the above fluences as inputs, we repeated the TID calculation for all three cases with results shown in Figure 16. Here the red curve is for the normal case (identical to the red curve in Figure 7). TIDs for Case 1 (the green curve) have larger values than the red for both ISS and TS2 orbits, and lower values for the two Transit orbits. This is consistent with the fluence curves as shown in Figures 12-15. The blue curve is for the Case 2 and all have lower values, and this curve is alike the 400% case in Figure 7. Note the ranges of TID in Figure 16 are almost enclosed by the ones in Figure 7, and therefore the general conclusions on the transmittance loss of coverglasses in Section 4 stay unchanged.

The vacancies are also calculated and shown in Figure 17, in which the red curve is for the normal case (identical to the red curve in Figure 10). Relative positions of the three curves are the same as those in Figure 16: the blue for Case 2 always have lower values than the normal case, while the green curve have two tipped-up ends with lower values for the two Transit orbits. Again the ranges of vacancies in Figure 16 are almost enclosed by the ones in Figure 10, and therefore the general conclusions on the displacement damages of solar PV cells in Section 5 stay unchanged.

Based on our calculations, we may go back to the observed power failure of Transit 4B satellite. From Figure 18, it is seen that Transit 4B solar cell shows a ~22% degradation over the 20 days after the detonation. Since the declining of current seems smooth over the 20 days with no visible abrupt changes, it is most likely that Sr+ ions only play a minor role in the radiation damages of Transit 4B solar cells compare to the MeV electrons.

Note this preliminary study aims to develop a protocol to estimate how HANE generated heavy ions affect satellite components. Here Sr$^+$ ions are used as the example with some assumed satellite solar panel geometry and materials. Any future experimental data of $Sr^+$ ions (or other heavy ions) can be very useful to constraint uncertainties in this study. Meanwhile, although the calculated results in this study may not be exact for a realistic solar panel design, the qualitative conclusions are valid. Finally, the methodology developed in this study can be easily extended to other particle species in the future.

# 7. Summary and Conclusions

Detonation of a nuclear bomb at high altitude creates a source of charged particles. The fission fragments after explosion are typically heavy ions including the radioactive strontium. The spatial and energy distributions of HANE-generated $Sr^+$ ions have been simulated by LANL's new RAM-HANE model, in which the effects of Coulomb collisions and charge exchanges neutralization is considered. It is shown that this artificial $Sr^+$ ion source can fill in the low altitude space, forming a short-lived ion artificial radiation belt that encircles the Earth, with a strong dependence on particles' energy, pitch angle and L-shell. Due to their very high NIEL values, these $Sr^+$ ions have extremely short ranges inside satellite materials and thus only affect the parts with no shielding/protection.

This study specifically focuses on the radiation effects of $Sr^+$ ions on satellite solar panels in LEO. Four different satellite orbits are selected for this study. $Sr^+$ ion fluences in these four orbits are fed to the MULASSIS code and SRIM model to quantify the damages. Specifically, the total ionizing doses are first calculated which have values of several Mrad for the low- and medium-apogee (100s to 1000 km) orbits. These TIDs suggest the transmittance losses of quartz coverglasses due to $Sr^+$ ions can be up to 6% or (much) lower. Therefore, we conclude that HANE-generated $Sr^+$ ions have limited darkening effects to solar cell coverglasses in LEO. In addition, displacement damages of $Sr^+$ ions on exposed GaAs cells are quantified by deriving their equivalent 1 MeV electron fluences, which have extremely high values between $10^{19} - 10^{20}$ per $cm^2$ in the low- and medium-apogee orbits. Hence, we conclude that HANE-generated $Sr^+$ ions may have severe or even fatal displacement damages to exposed solar PV cells (e.g., with gaps in top layers) on satellites in LEO. The above radiation effects of $Sr^+$ ions are much less significant for the orbits with high apogees beyond ten thousand kilometers. Parameter sensitivity studies on the charge exchange cross-sections, atmosphere density profiles, and burst local time positions have been conducted with the above conclusions unaltered. We expect the methodology developed in this study can be extended to other particle species in the future.

# Acknowledgements

This work is supported at LANL under the LDRD Mission Foundations Research program (project 20190528ER). We acknowledge the SPENVIS website for the use of MULASSIS code, and also thanks to the SRIM team for making the software available for public use.  We are also grateful to Jeffrey S. George at LANL who has provided insightful suggestions on comparing SRIM and Geant4 results.

# Tables

**Table 1. Orbit parameters and periods of selected satellites**

| # | Satellite (abbrev.) | Perigee (km) | Apogee (km) | Inclination (deg) | Period (min) |
|---|---|---|---|---|---|
| 1 | Inter. Space Station (ISS) | 408 | 410 | 51.6 | 93 |
| 2 | Transit 4A (TR4A) | 881 | 998 | 66.8 | 104 |
| 3 | Transit 4B (TR4B) | 956 | 1106 | 32.4 | 106 |
| 4 | Telstar 2 (TS2) | 972 | 10,803 | 42.7 | 225 |

# Figures

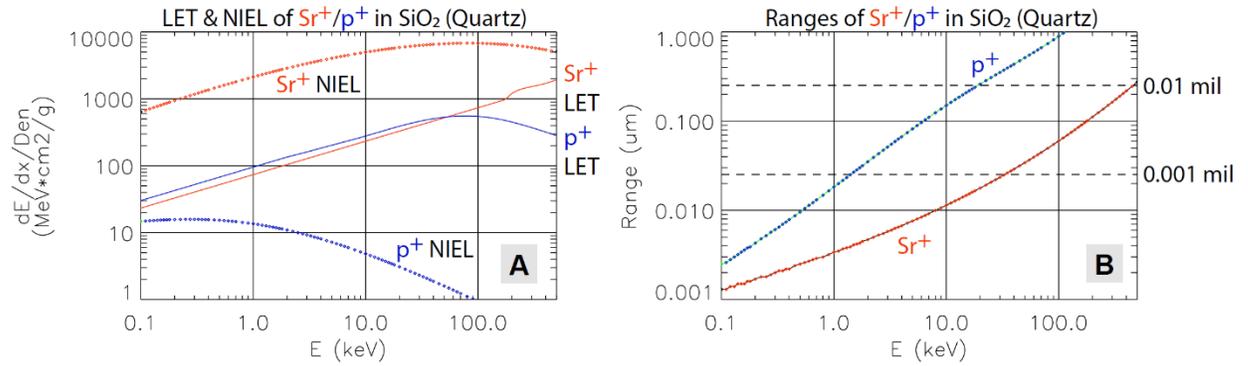

**Figure 1 Stopping power curves and ranges of Sr⁺ ions in quartz (SiO₂). A)** Linear energy transfer (LET, solid line in red) and non-ionizing energy loss (dashed red) curves for Si⁺ ions as a function of particle energy in SiO₂ (quartz, ICRU-245), compared to those of protons (in blue). **B)** Range curve of Sr⁺ ions (red) in quartz compared to protons (blue). Ranges of 0.001 and 0.01 mil are marked out by horizontal dashed lines.

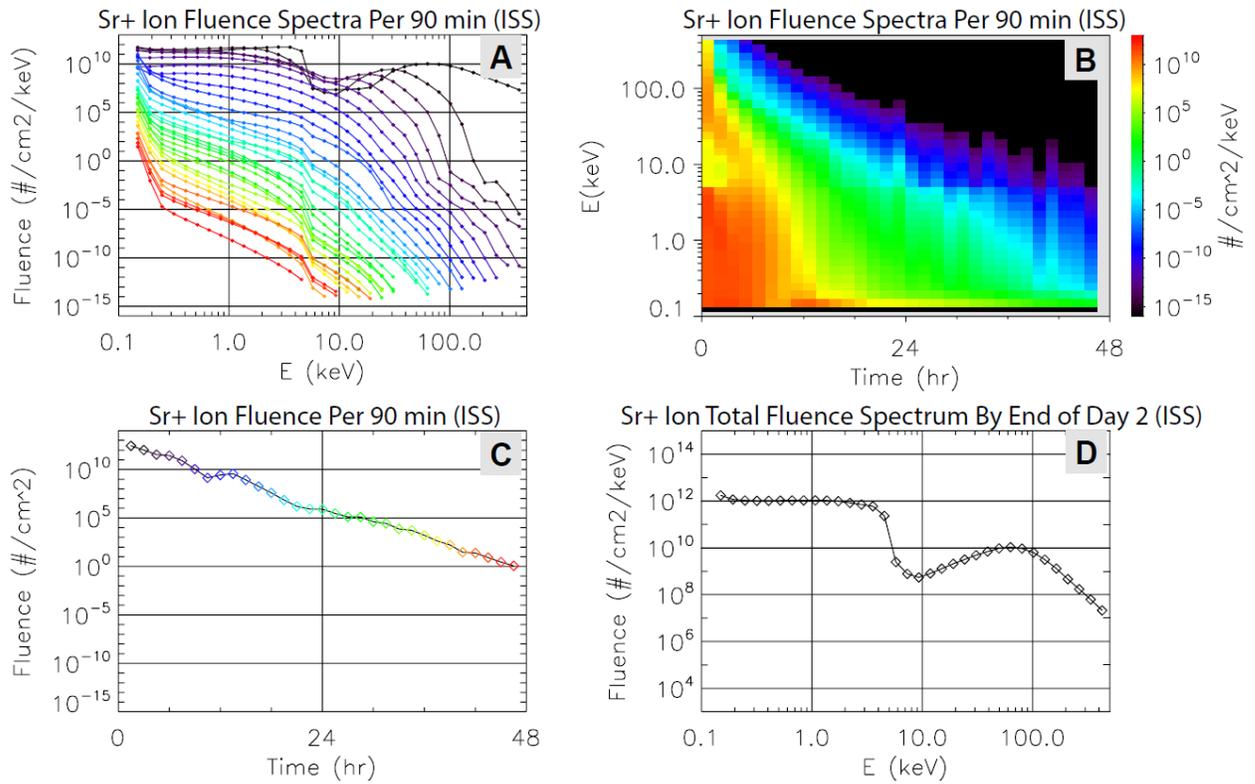

**Figure 2 Decays of Sr⁺ ions over two days in the ISS orbit. A)** Energy differential Sr$^+$ ion fluence in each 90 min time bin as observed along ISS orbit. Different color indicates different time as in Panel C. **B)** Sr$^+$ ion spectra in all time bins are replotted to highlight the decay speeds varying with different energies. **C)** Energy-integrated ion fluences decrease quickly over time. Different colors correspond to different time bins. **D)** Total ion fluence accumulated over two days after the detonation is plotted as a function of ion energy.

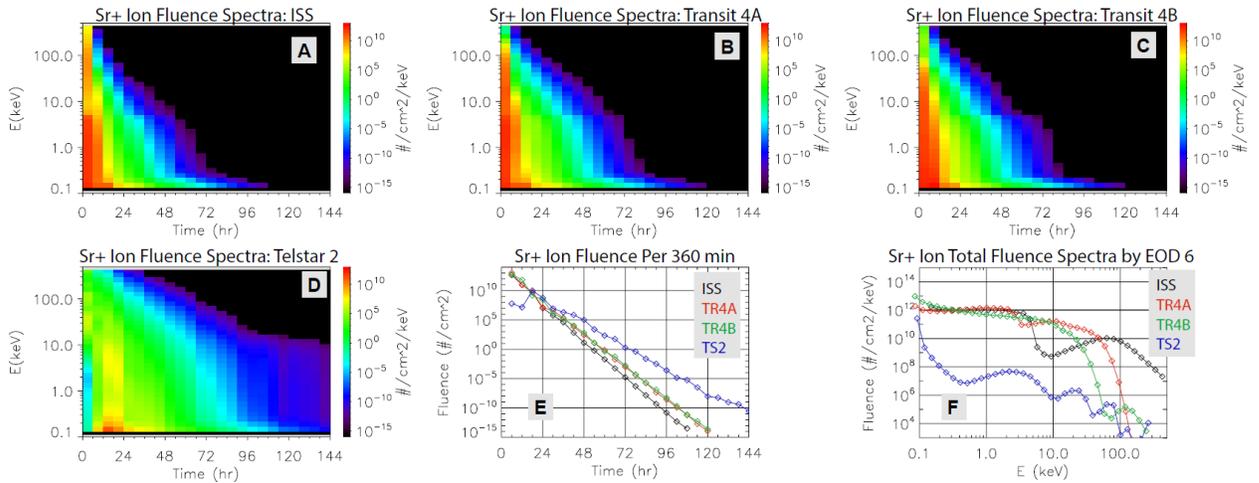

**Figure 3 Distributions of Sr⁺ ions over six days experienced in four selected orbits. A to D)** Energy-differential ion fluences decay over time along the orbits of ISS, Transit 4A, Transit 4B and Telstar 2 over six days after the detonation. **E)** Temporal decays of energy-integrated ion fluences in four satellite orbits over six days. **F)** Total ion fluences accumulated over six days are plotted as a function of energy for the four orbits.

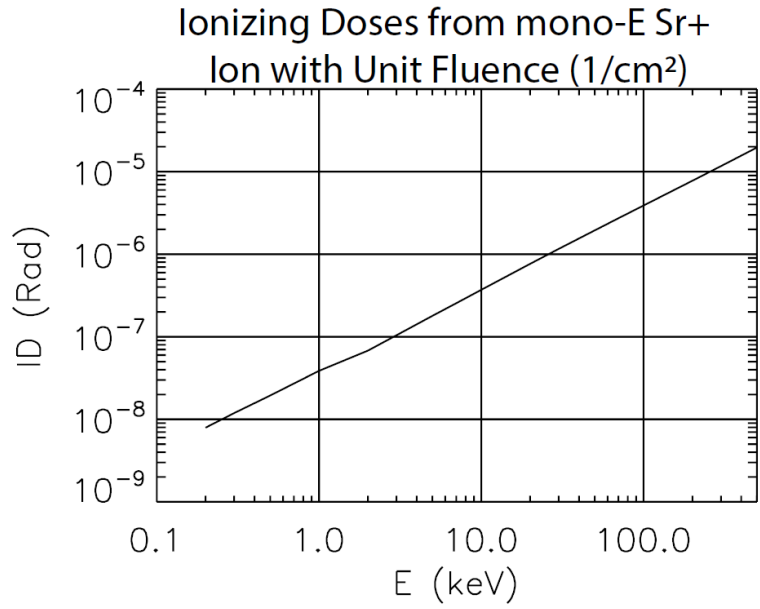

**Figure 4 Ionizing doses on the quartz slab (with a thickness of 0.015 mil) contributed from incident mono-energy Sr$^+$ ions with a unit fluence of 1/cm$^2$.**

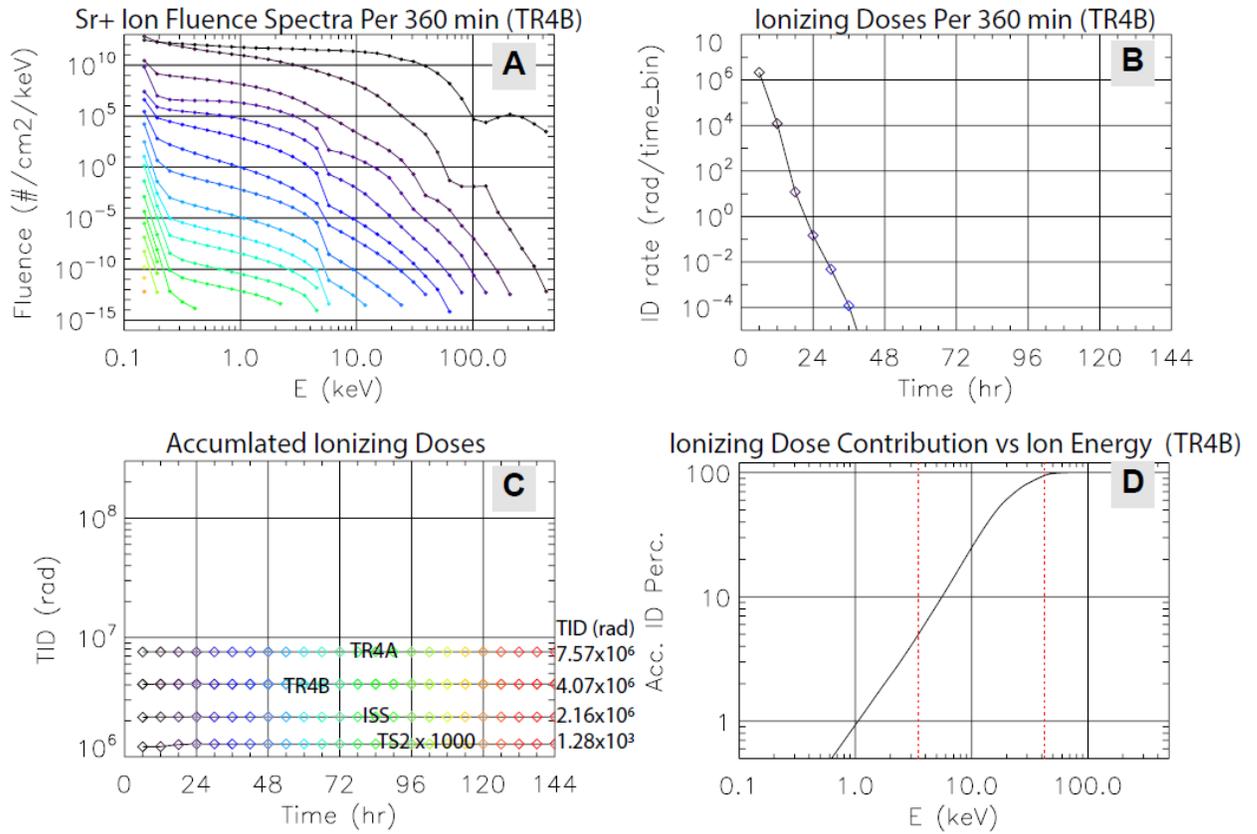

**Figure 5 Input Sr⁺ ion spectra and calculated ionizing doses to the quartz coverglass of satellite solar cells. A)** Example energy-differential Sr⁺ ion fluences for each 360-min time bin experienced by Transit 4B, are the inputs for ionizing dose calculations. Data points in Panels A, B and C are color-coded for time in the same way. **B)** The ionizing dose for each time bin is plotted and color-coded for time. **C)** Ionizing doses are accumulated over time for all four satellite orbits, and the total ionizing dose values after six days are presented to the right. **D)** Effective energy range (between the two red vertical dashed lines) is determined from the normalized accumulative dose percentage curve for satellite Transit 4B.

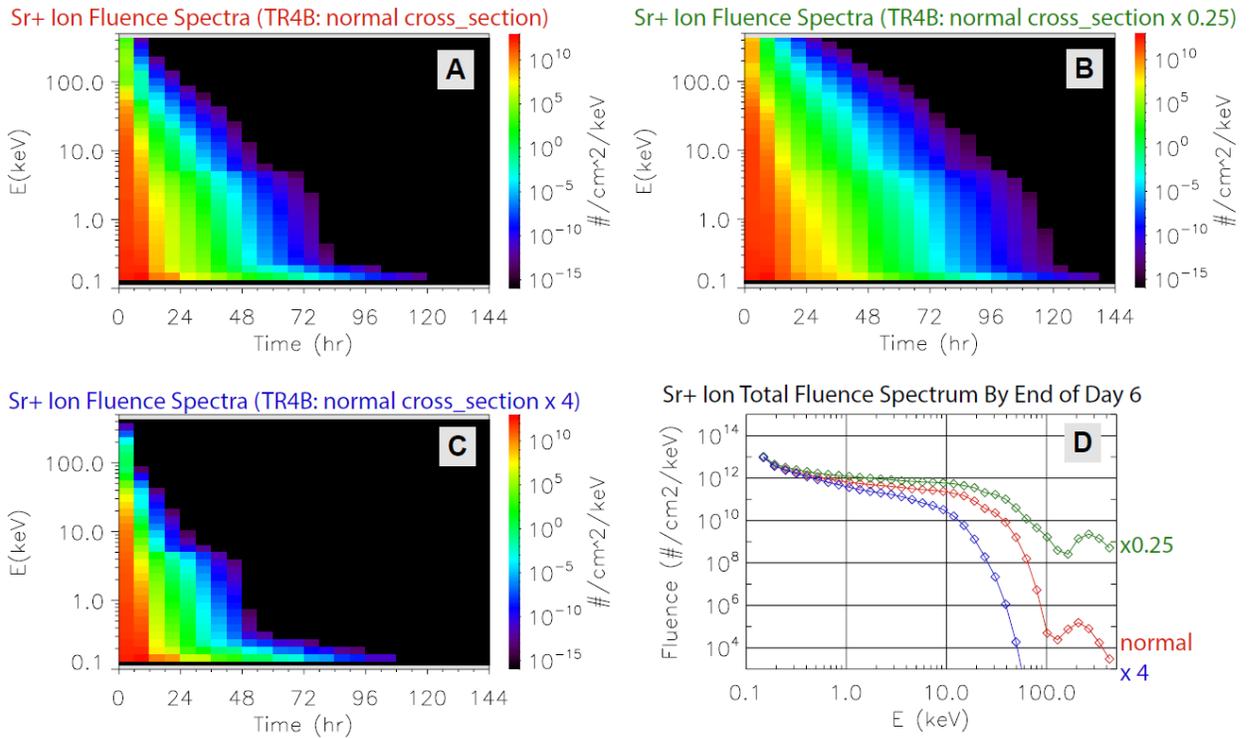

**Figure 6 Charge-exchange cross sections significantly impact the decay of Sr⁺ ions.** **A)** Distributions of Sr$^+$ ions experienced in Transit 4B orbit decay over time with "normal" charge exchange cross sections. **B)** Distributions of Sr$^+$ ions decay more slowly with the charge exchange cross sections reduced by a factor of four. **C)** Distributions of Sr$^+$ ions decay more rapidly with the charge exchange cross-sections increased by a factor of four. **D)** Total ion fluences accumulated over six days after the detonation are plotted as a function of energy in TR4B orbit for three cases: red for "normal" cross sections, green for reduced cross section, and blue for increased cross sections.

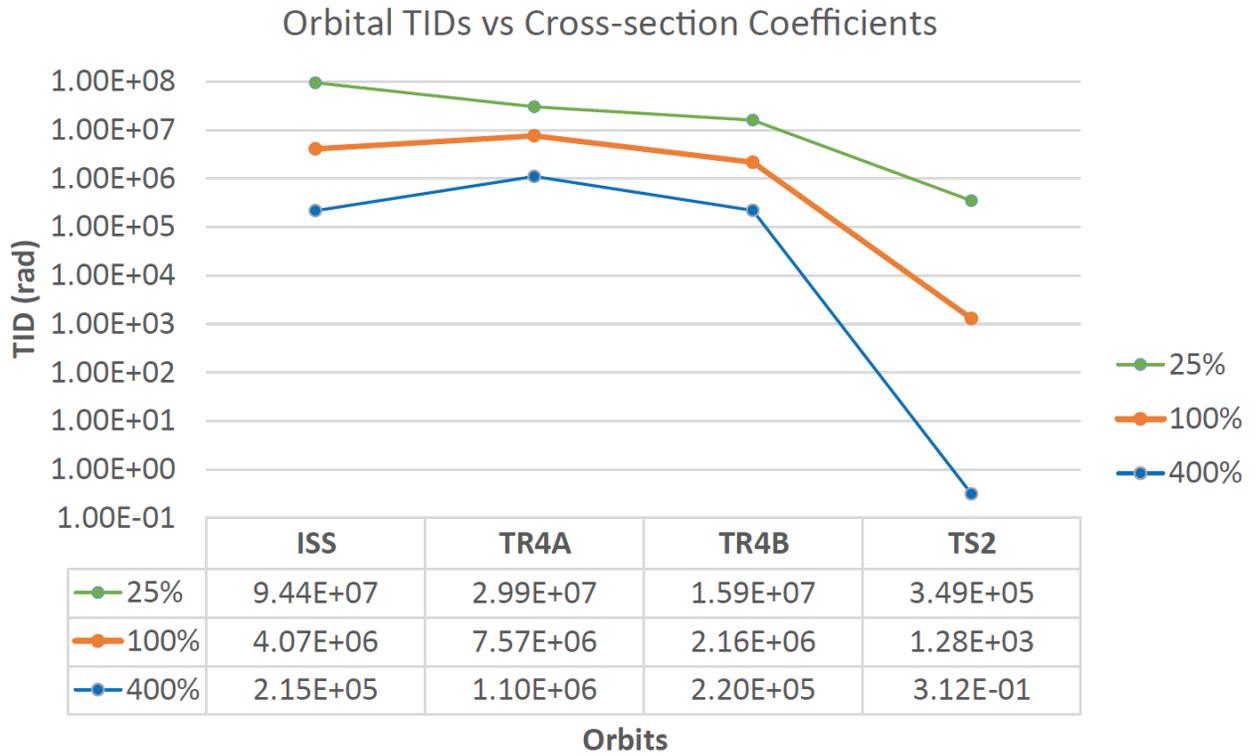

| | ISS | TR4A | TR4B | TS2 |
|---|---|---|---|---|
| 25% | 9.44E+07 | 2.99E+07 | 1.59E+07 | 3.49E+05 |
| 100% | 4.07E+06 | 7.57E+06 | 2.16E+06 | 1.28E+03 |
| 400% | 2.15E+05 | 1.10E+06 | 2.20E+05 | 3.12E-01 |

**Figure 7 TIDs for all four selected satellite orbits with three different charge exchange cross-sections.** Orange curve is for the normal cross-sections, and the green/blue curve is for the decreased/increased cross-sections by a factor of 4.

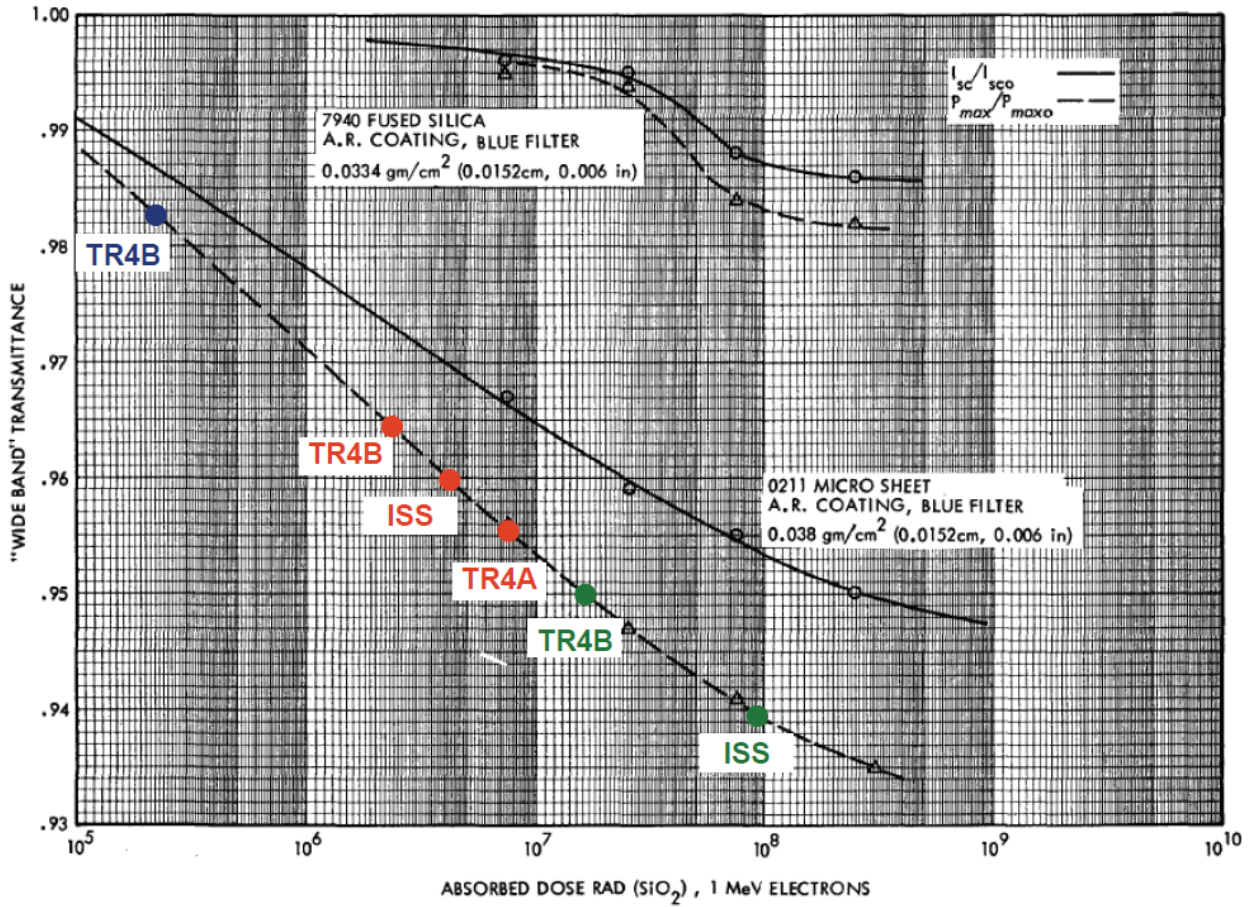

**Figure 8 Transmittance of quartz solar-cell coverglass reduced as a function of Sr+ ion TID.** The bottom two curves are for 0211 micro sheet coating with a thickness of 0.0152 cm, and the top two curves are for 7940 fused silica. Red symbols on the micro sheet power curve are for three satellites with "normal" charge exchange cross sections, green (blue) for Transit 4B orbit with reduced (increased) cross sections. Based on Carter and Tada [1973].

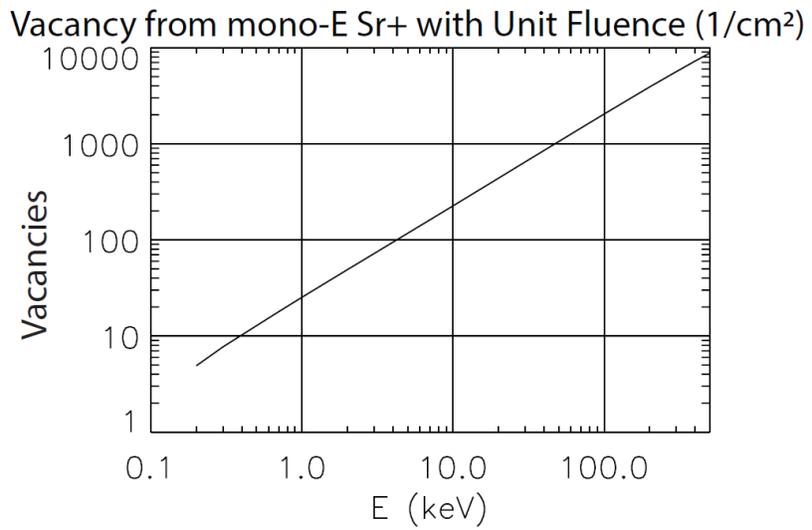

**Figure 9 Vacancies are introduced inside a GaAs slab (with a thickness of 0.025 mil) from incident mono-energy Sr⁺ ions with unit fluence of 1/cm².**

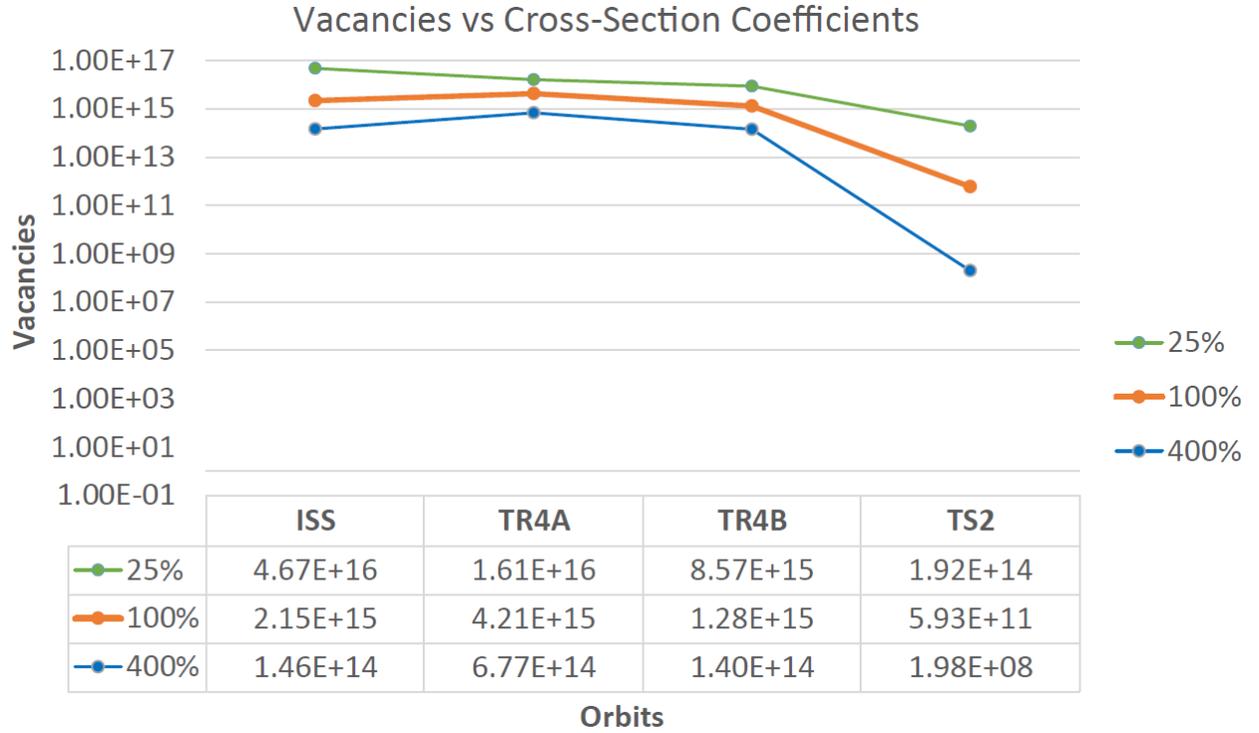

| | ISS | TR4A | TR4B | TS2 |
|---|---|---|---|---|
| ●— 25% | 4.67E+16 | 1.61E+16 | 8.57E+15 | 1.92E+14 |
| ●— 100% | 2.15E+15 | 4.21E+15 | 1.28E+15 | 5.93E+11 |
| ●— 400% | 1.46E+14 | 6.77E+14 | 1.40E+14 | 1.98E+08 |

**Figure 10 Vacancies in the given GaAs solar cell for all four selected satellite orbits with three different charge exchange cross sections.** Orange curve is for the normal cross-sections, and the green/blue curve is for the decreased/increased cross-sections by a factor of 4.

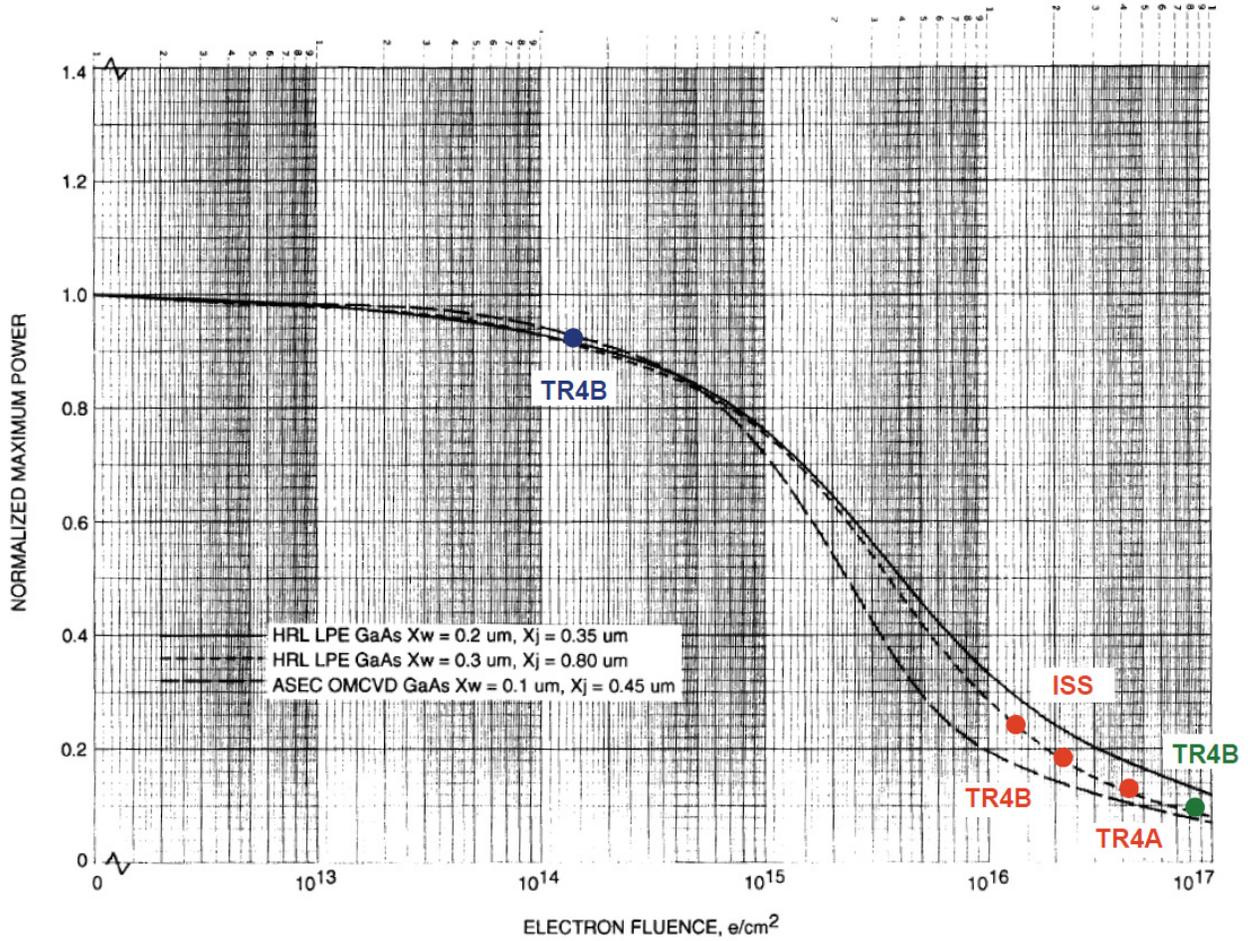

**Figure 11 GaAs solar cell output power decreases as a function the equivalent 1 MeV electrons from Sr+ ions.** Red symbols mark the equivalent 1 MeV electron fluences reduced by a factor of 3.33 x 10³ for three satellites with "normal" charge exchange cross sections, green (blue) for Transit 4B orbit with reduced (increased) cross sections. The $x_w$ is cell's window thickness and $x_j$ is junction depth. Modified from [Anspaugh, 1996].

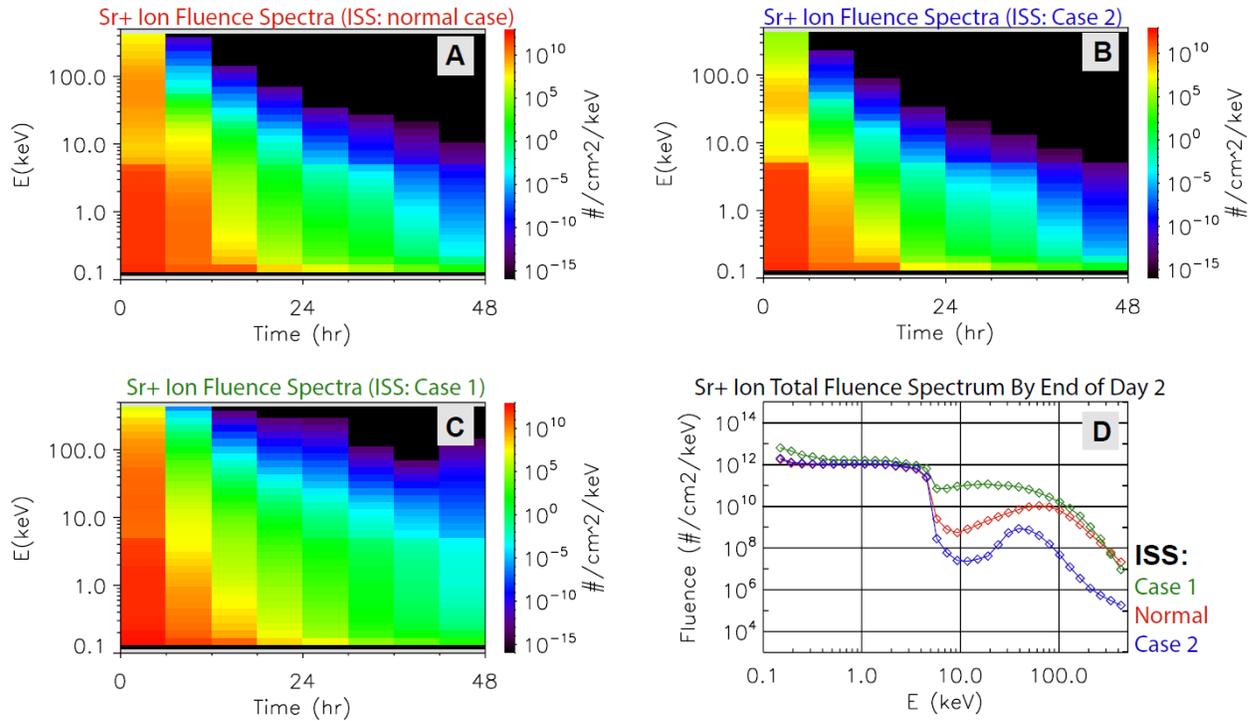

**Figure 12 Model parameters affect the total Sr+ ions experienced in ISS orbit.  A)** Distributions of Sr$^+$ ions in ISS orbit decay over time in the "normal" case with the burst at midnight. **B)** Distributions of Sr+ ions decay more quickly in Case 2 with the burst at midnight and modified atmosphere density profiles. **C)** Distributions of Sr+ ions decay more slowly in Case 1 with the burst moved to noon and modified atmosphere density profiles. **D)** Total ion fluences accumulated over two days after detonation are plotted as a function of energy for three cases: red for the "normal" case, green for Case 1, and blue for Case 2.

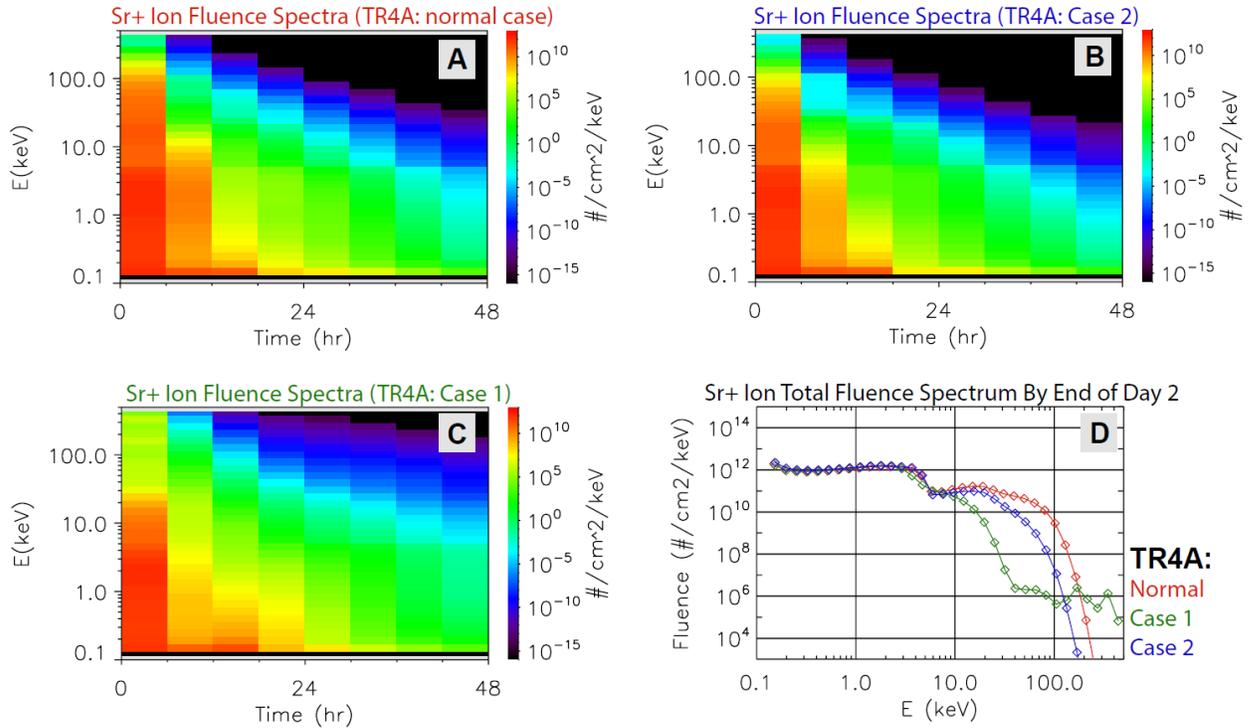

**Figure 13 Model parameters affect the total Sr⁺ ions experienced in Transit 4A orbit.** Panels in the same format as Figure 12. In panel D, total ion fluences accumulated over two days after detonation are plotted as a function of energy for three cases: red for the "normal" case, green for Case 1, and blue for Case 2.

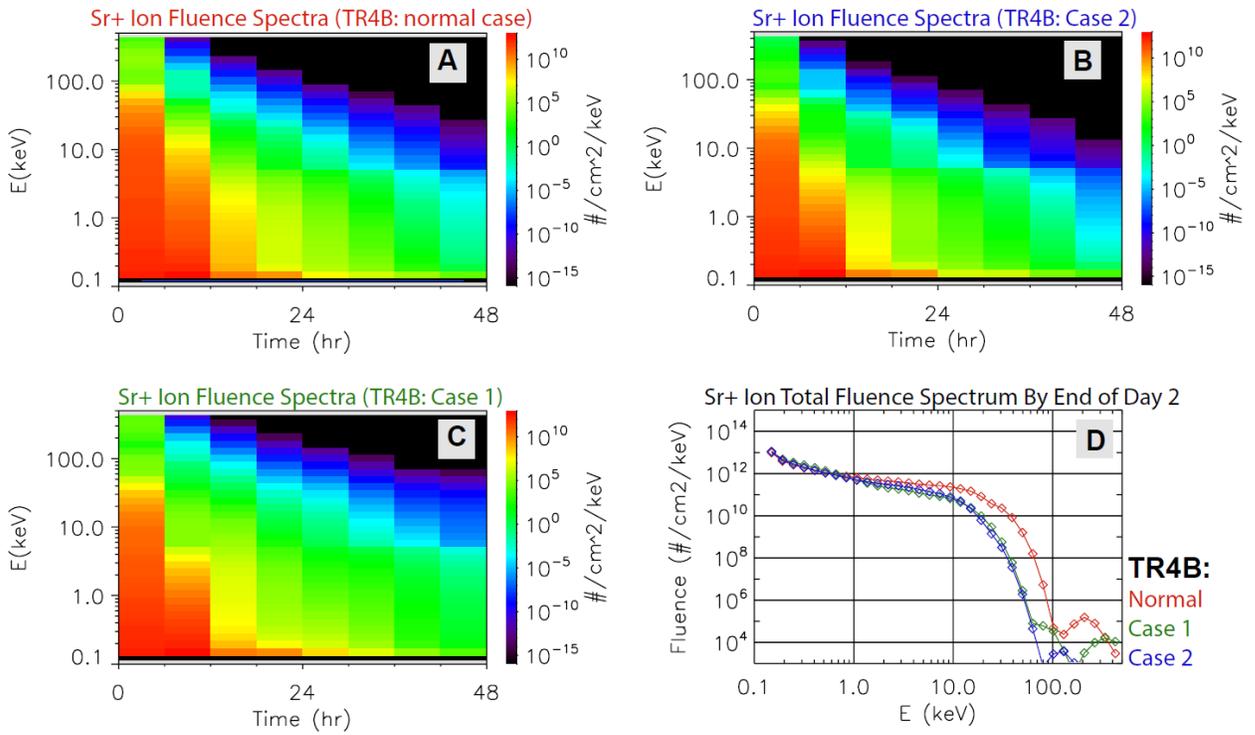

**Figure 14 Model parameters affect the total Sr+ ions experienced in Transit 4B orbit.** Panels in the same format as Figure 12. In panel D, total ion fluences accumulated over two days after detonation are plotted as a function of energy for three cases: red for the "normal" case, green for Case 1, and blue for Case 2.

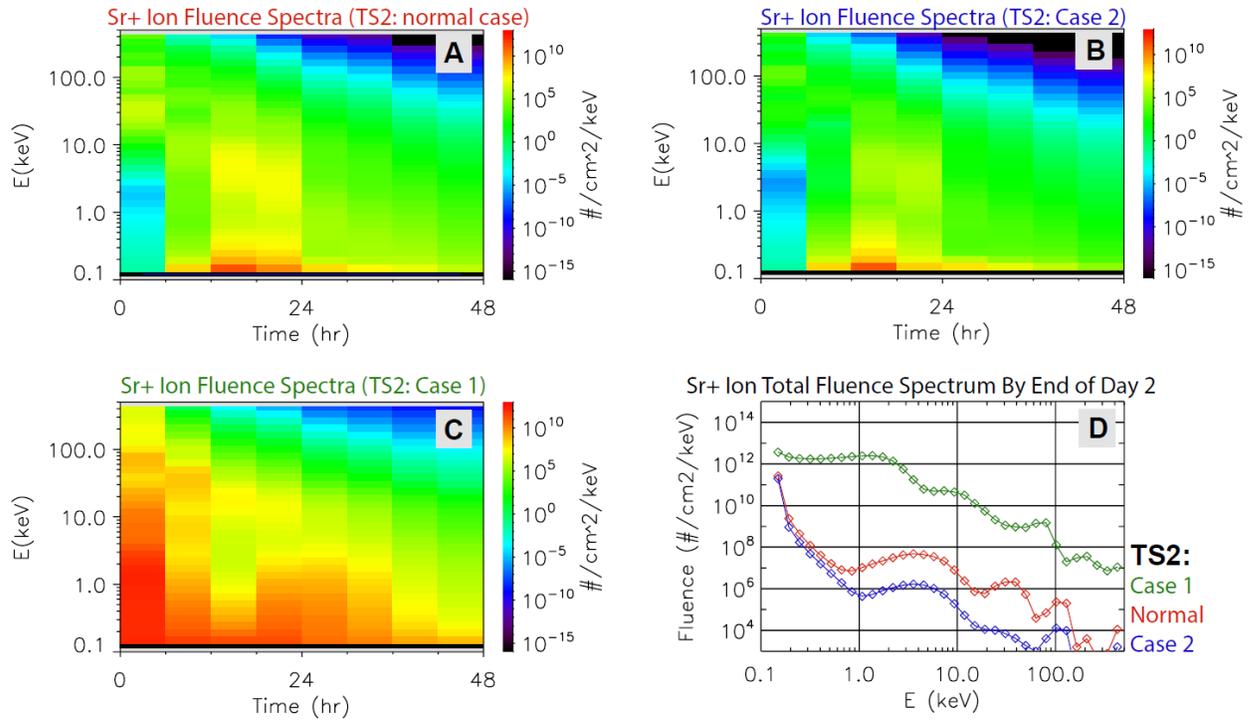

**Figure 15 Model parameters affect the total Sr+ ions experienced in Telstar 2 orbit.**
Panels in the same format as Figure 12. In panel D, total ion fluences accumulated over
two days after detonation are plotted as a function of energy for three cases: red for the
"normal" case, green for Case 1, and blue for Case 2.

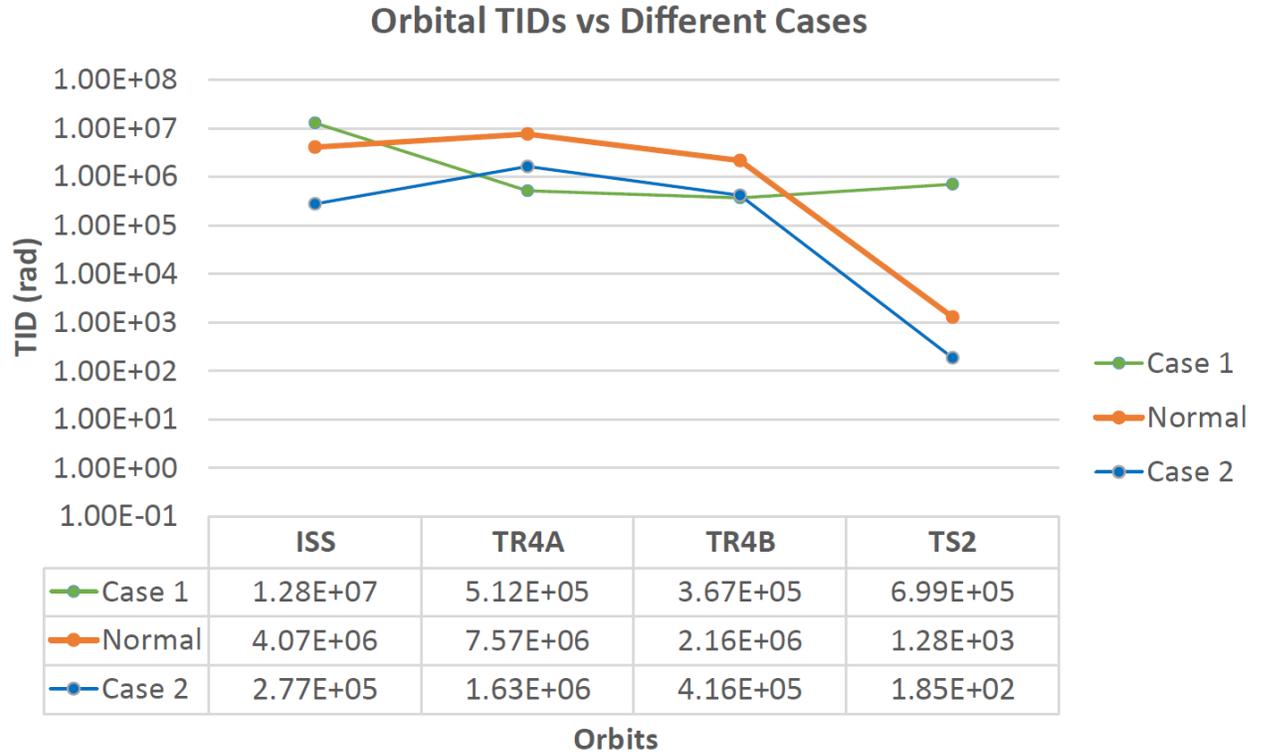

| | ISS | TR4A | TR4B | TS2 |
|---|---|---|---|---|
| Case 1 | 1.28E+07 | 5.12E+05 | 3.67E+05 | 6.99E+05 |
| Normal | 4.07E+06 | 7.57E+06 | 2.16E+06 | 1.28E+03 |
| Case 2 | 2.77E+05 | 1.63E+06 | 4.16E+05 | 1.85E+02 |

**Figure 16 TID for all four selected satellite orbits in three different cases as defined in the text.**

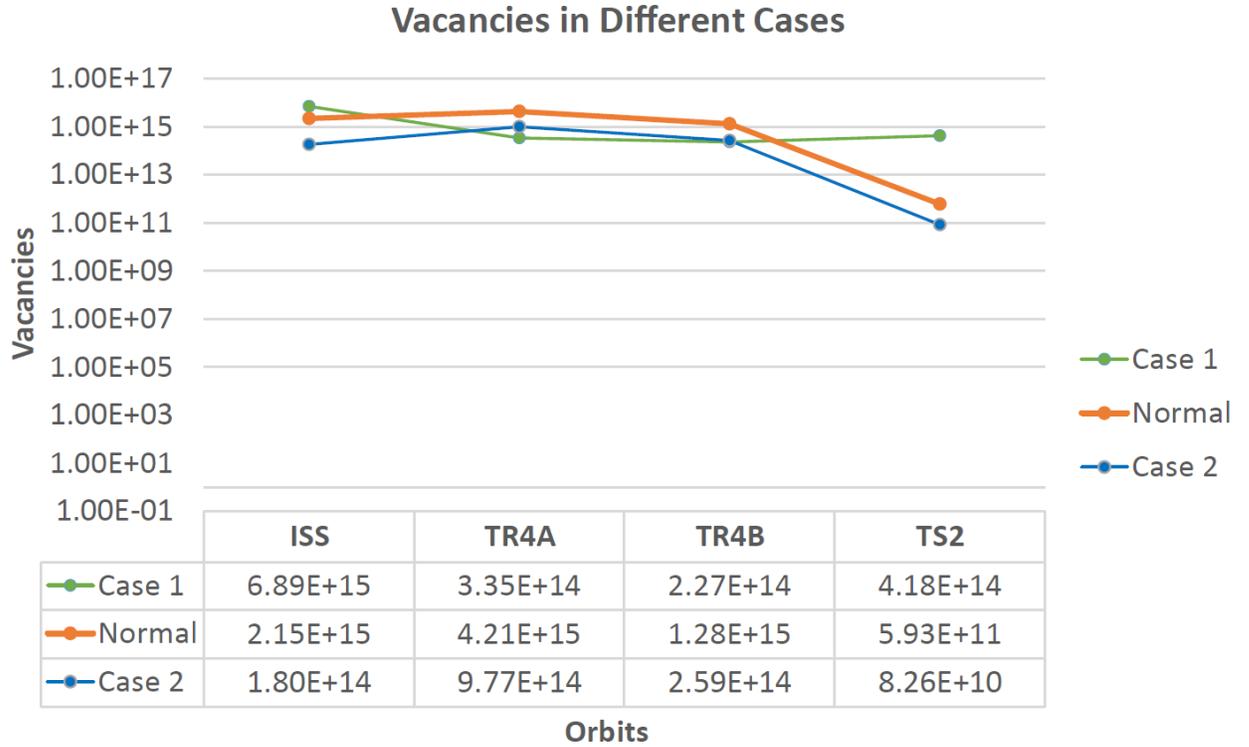

**Figure 17 Vacancies in satellite GaAs solar cells for all four selected satellite orbits in three different cases.**

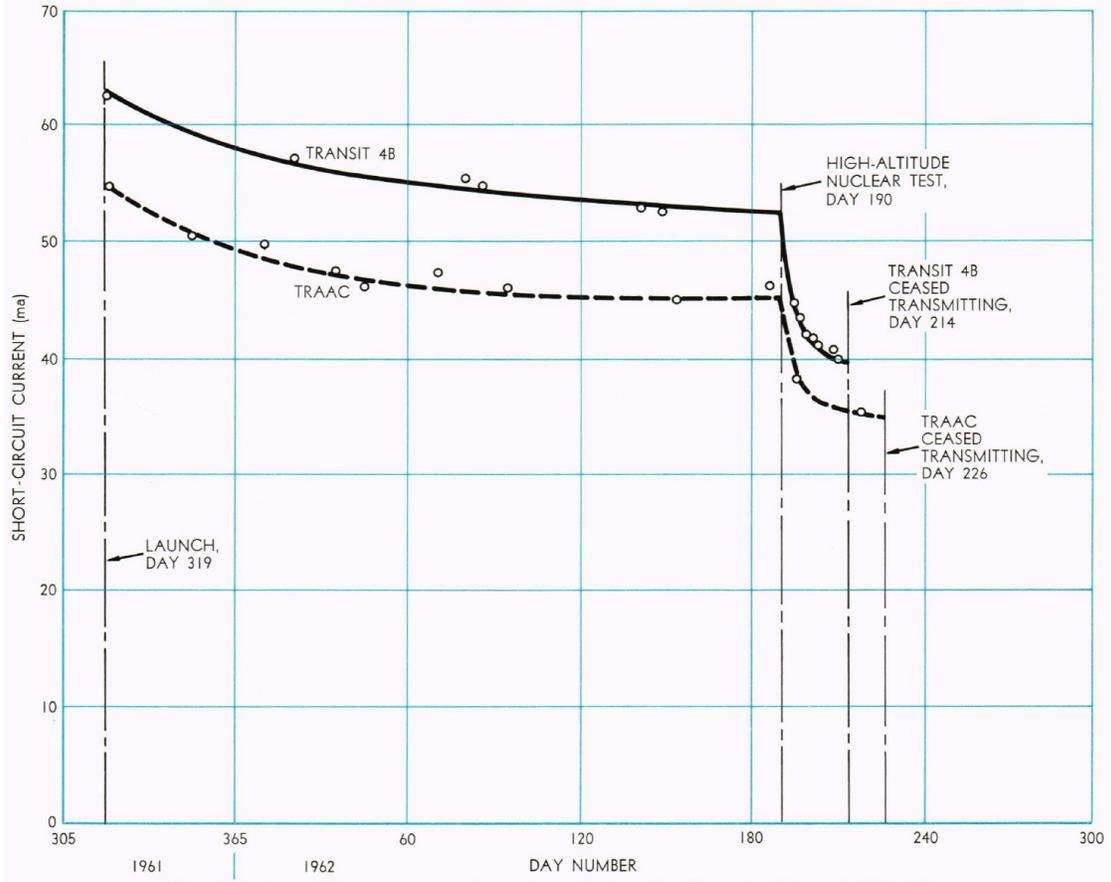

**Figure 18. Solar cell outputs decline dramatically after Starfish on day 290 for Transit 4B (solid) and TRAAC (dashed).** Copied from [Fischell, 1962].

## Appendix Figures

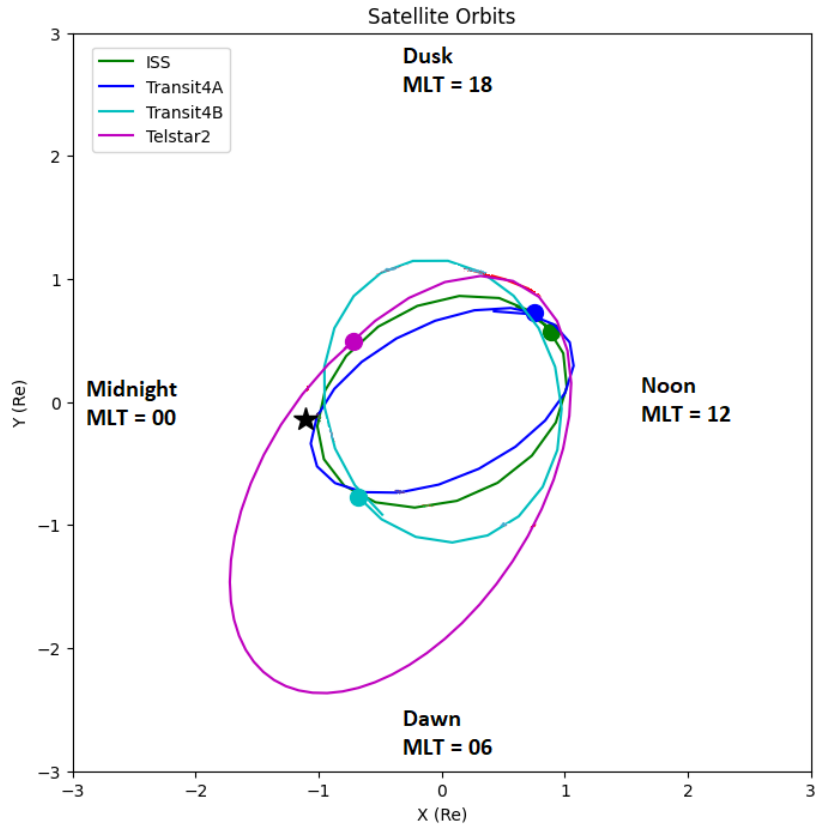

**Figure A1. Top view of satellite orbits summarized in Table 1.** Here Starfish burst location (the black star symbol) is at midnight, and the dots indicate satellite positions at the burst time.

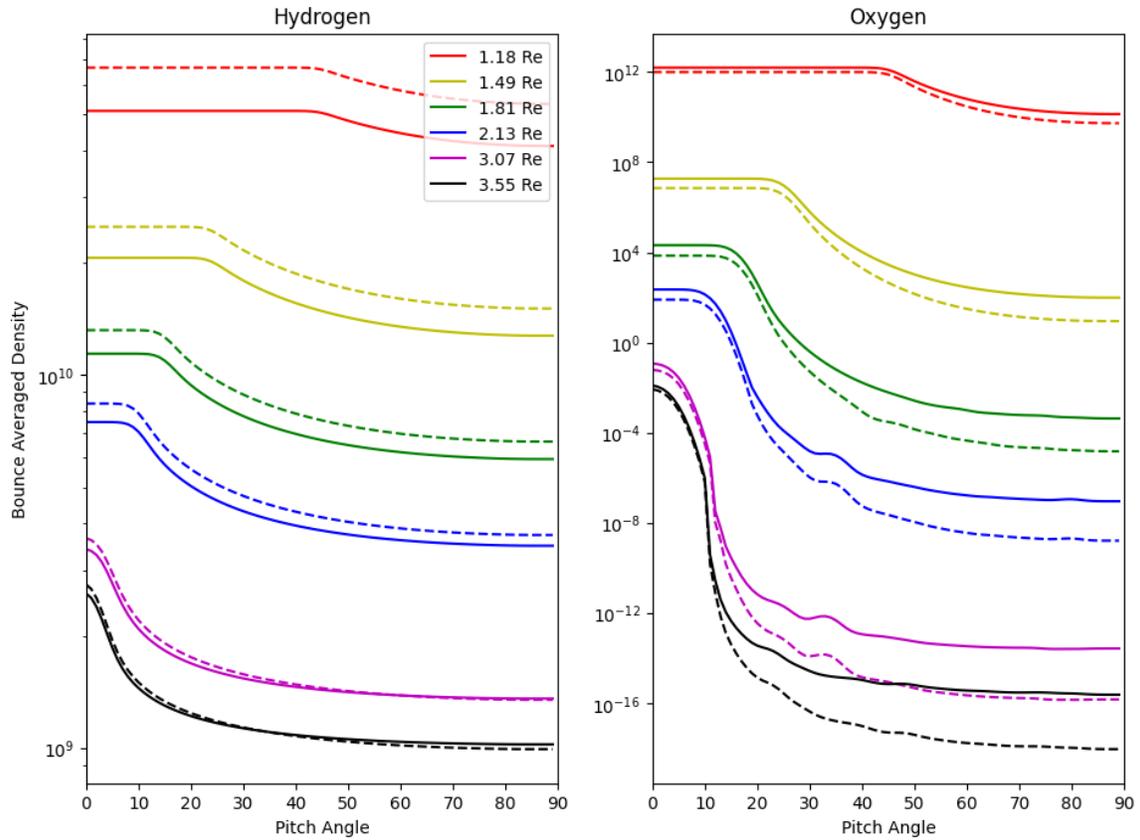

**Figure A2 Atmosphere densities used by RAM-HANE simulations.** Accumulated H (left panel) and O (right panel) densities encounter by Sr+ ions in their each bounce movement are plotted as a function of particles' pitch angle at different L-shells (different colors). These curves are for the midnight (MLT = 0) position. Solid lines are for the normal case, and dashed lines are for Cases 1 and 2.